\documentclass[12pt,numbers,sort&compress]{article}

\usepackage{natbib}
\usepackage{amsmath}
\usepackage{amsthm}
\usepackage{amsfonts}
\usepackage{amssymb}
\usepackage[all]{xy}
\usepackage{graphicx}
\usepackage{color}
\usepackage[mathscr]{eucal}
\usepackage[colorlinks=true,linktocpage]{hyperref}
\usepackage[isu,bf]{caption}
\newlength{\xtrawidth}
\newlength{\xtraheight}
\newcommand{\Rep}[1]{\ensuremath{\mathbf{\underline{#1}}}}
\newcommand{\barRep}[1]{\ensuremath{\overline{\Rep{#1}}}}
\newcommand{\detinv}{\operatorname{det}^{-1}}
\DeclareMathOperator{\Ext}{Ext}
\newcommand{\Z}{\mathbb{Z}}
\newcommand{\C}{\mathbb{C}}
\newcommand{\CP}{\mathbb{P}}
\DeclareMathOperator{\Span}{span}
\def\dP9{\ensuremath{\mathrm{dP}_9}}
\DeclareMathOperator{\Hsupp}{Hsupp}

\DeclareMathOperator{\Tr}{Tr}
\DeclareMathOperator{\ad}{ad}

\newcommand{\itemlefteq}[1]{\makebox{\rlap{#1}}}
\newenvironment{itemequation}{%
\begin{minipage}[b]{\linewidth}\begin{equation}}{%
\end{equation}\end{minipage}}

\begin{document}
\begin{titlepage}
  \begin{flushright}
    hep-th/0606166
  \end{flushright}
  \vspace*{\stretch{1}}
  \begin{center}
     \LARGE 
     Dynamical SUSY Breaking in Heterotic M-Theory
  \end{center}
  \vspace*{\stretch{2}}
  \begin{center}
    \begin{minipage}{\textwidth}
      \begin{center}
        \large         
        Volker Braun$^{1,2}$, 
        Evgeny I. Buchbinder$^{3}$,
        Burt A.~Ovrut$^1$
      \end{center}
    \end{minipage}
  \end{center}
  \vspace*{1mm}
  \begin{center}
    \begin{minipage}{\textwidth}
      \begin{center}
        ${}^1$ Department of Physics,
        ${}^2$ Department of Mathematics
        \\
        University of Pennsylvania,        
        Philadelphia, PA 19104--6395, USA
      \end{center}
      \begin{center}
        ${}^3$ School of Natural Sciences, 
        Institute for Advanced Study\\
        Einstein Drive, Princeton, NJ 08540
      \end{center}
    \end{minipage}
  \end{center}
  \vspace*{\stretch{1}}
  \begin{abstract}
    \normalsize 
    It is shown that four-dimensional $N=1$ supersymmetric QCD with
    massive flavors in the fundamental representation of the gauge
    group can be realized in the hidden sector of $E_{8} \times E_{8}$
    heterotic string vacua. The number of flavors can be chosen to lie
    in the range of validity of the free-magnetic dual, using which
    one can demonstrate the existence of long-lived meta-stable
    non-supersymmetric vacua. This is shown explicitly for the gauge
    group $Spin(10)$, but the methods are applicable to $Spin(N_{c})$,
    $SU(N_{c})$ and $Sp(N_{c})$ for a wide range of color index
    $N_{c}$. Hidden sectors of this type can potentially be used as a
    mechanism to break supersymmetry within the context of heterotic
    M-theory.
  \end{abstract}
  \vspace*{\stretch{5}}
  \begin{minipage}{\textwidth}
    \underline{\hspace{5cm}}
    \centering
    \\
    Email: 
    \texttt{vbraun, ovrut@physics.upenn.edu},
    \texttt{evgeny@sns.ias.edu}.
  \end{minipage}
\end{titlepage}

\section*{Introduction}

It has long been known~\cite{1} that a four-dimensional $N=1$ super
Yang-Mills theory with purely massive vector-like matter has a
non-vanishing Witten index and that this implies the existence of
supersymmetry preserving vacua. For a super Yang-Mills theories to
have no supersymmetric vacua at all, it must either be
chiral~\cite{2,3} or, if non-chiral, it must have massless
matter~\cite{4,5}.  Theories satisfying these requirements are
complicated and their incorporation into realistic particle physics
models has been difficult to achieve.

Recent results in various contexts in string theory~\cite{6} have
stimulated interest in four-dimensional $N=1$ theories with
\emph{both} supersymmetric and non-supersymmetric vacua, the
supersymmetry breaking states being meta-stable and long-lived. With
this in mind, the authors of~\cite{7} have re-examined
four-dimensional $N=1$ SQCD theories with purely massive vector-like
matter. Using the free-magnetic dual description of the theory in the
infrared, they can determine the vacuum structure of the strongly
coupled gauge theory. They find that, in addition to the requisite
supersymmetric vacua, there exist meta-stable supersymmetry breaking
vacua that, under the appropriate circumstances, can be long-lived.

Specifically, they show the following. They first consider $N=1$ 
supersymmetric QCD with gauge group $SU(N_{c})$ and $N_{f}$ massive
flavors in the fundamental representation. Taking $N_{f}$ to be in the
free-magnetic range, $N_{c}+1 \leq N_{f} < \frac{3}{2}N_{c}$, they prove
that, in addition to the $N_{c}$ expected supersymmetric vacua, there exist
supersymmetry breaking meta-stable vacua. Furthermore, in the limit that
\begin{equation}
  |\epsilon|= \sqrt{\left|\frac{m}{\Lambda}\right|} \ll 1,
  \label{1}
\end{equation}
where $m$ is the typical scale of the quark masses and $\Lambda$ is
the strong-coupling scale, the meta-stable vacua are long-lived.  In
particular, as the parameter $|\epsilon| \rightarrow 0$, the life-time
of these non-supersymmetric states can easily exceed the age of the
Universe.

The authors of~\cite{7} then generalized their results to gauge groups
$Spin(N_{c})$ and $Sp(N_{c})$. In this paper, we will be particularly
interested in the $Spin(N_{c})$ case. Consider $N=1$ supersymmetric
QCD with gauge group $Spin(N_{c})$ and $N_{f}$ massive flavors in the
fundamental representation. Then, in the free-magnetic range (including
two special cases),
\begin{equation}
  N_{c}-4 ~\leq~ N_{f} ~<~ \frac{3}{2}(N_{c}-2),
  \label{2}
\end{equation}
it was shown in~\cite{7} that, in addition to the $N_{c}-2$ expected
supersymmetric vacua, there are supersymmetry breaking meta-stable vacua
which, when inequality~\eqref{1} is satisfied, are long-lived.

These results open the door to a re-examination of dynamical
supersymmetry breaking in many contexts. In particular, it is of
interest to ask whether four-dimensional $N=1$ supersymmetric QCD,
with a matter spectrum consisting of massive vector-like fundamental
representations only and for which the number of flavors lies in the
free-magnetic range, can be found as the low energy theory of a string
compactification. Furthermore, can such an effective theory be
incorporated into a realistic string theory of particle physics? 
Type II string theory realizations of dynamical supersymmetry
breaking were recently obtained in~\cite{Franco, Ooguri}.
In this paper, we will show that the answer to both questions is
also affirmative in the context of heterotic M-theory.
Specifically, we prove the following. Consider the $E_{8}
\times E_{8}$ heterotic superstring, either for weak or strong string
coupling, and focus on the hidden sector gauge group $E_{8}$. We will
present an explicit elliptically fibered Calabi-Yau threefold $X$ and
a slope-stable holomorphic vector bundle $V$ with structure group
$SU(4)$ that lead to an $N=1$ supersymmetric $Spin(N_{c})$, $N_{c}=10$
theory of QCD in four-dimensions. The matter spectrum of this theory
consists entirely of $N_{f}$ massive 
fundamental \Rep{10}
representations, where $N_f$ can take any integer value. Noting that
for $N_{c}=10$ the constraint eq.~\eqref{2} becomes
\begin{equation}
  6 \leq N_{f} < 12,
  \label{3}
\end{equation}
we conclude that this hidden sector is a heterotic string vacuum with
eight supersymmetric vacua and meta-stable non-supersymmetric vacua.
Furthermore, the typical mass parameter $m$ is a linear function of
the vector bundle moduli whose vacuum expectation values can be
adjusted to satisfy eq.~\eqref{1}. Hence, the supersymmetry breaking
vacua are long-lived and of phenomenological interest. Finally, a
hidden sector of this type can, in principal, be embedded in heterotic
$M$-theory~\cite{8} models of the type discussed in~\cite{9}. This
could provide the mechanism, or one of several mechanisms, for
breaking supersymmetry in realistic heterotic string
compactifications.
The technical details leading to these results will appear 
in~\cite{13}.

To summarize, in this paper we present a hidden sector whose low
energy theory is $N=1$ SQCD with gauge group $Spin(10)$ and show that
this admits meta-stable, long-lived non-supersymmetric vacua.  This
gauge group, and the slope-stable holomorphic vector bundle that leads
to it, were chosen for mathematical convenience. It is straightforward
to generalize this to vector bundles whose low energy theories are
$N=1$ SQCD with gauge groups $SU(N_{c})$, $Spin(N_{c})$ and
$Sp(N_{c})$, for a range of values of $N_{c}$, and $N_{f}$ massive
vector-like fundamental representations where $N_{f}$ is in the range
specified in~\cite{7}. These hidden sectors will also admit
meta-stable vacua which are long-lived. Such generalized hidden
sectors can be used to break supersymmetry in heterotic
$M$-theory particle physics models. These generalized theories will be
presented elsewhere.

\section*{The Heterotic Vacua}

\subsection*{The Calabi-Yau Threefold}

We construct our hidden sector by compactifying the $E_{8} \times
E_{8}$ heterotic string on a simply-connected Calabi-Yau threefold $X$
which is elliptically fibered over the complex surface \dP9. It was
shown in~\cite{10} that $X$ must factorize into the fibered product
$X=B_{1} \times_{\CP^1} B_{2}$, where $B_{1}$ and $B_{2}$ are both
\dP9 surfaces. Either can be thought of as the base space and the
projection maps are denoted by $\pi_{i}:X \rightarrow B_{i}$ with
$i=1,2$ respectively. Each \dP9 surface is itself elliptically fibered
over the the identified projective sphere $\CP^1$, with the projection
mappings $\beta_{i}:B_{i} \rightarrow \CP^1$ for $i=1,2$.  It is
straight forward to show that the number of K\"ahler and complex
structure moduli of $X$ is given by $h^{1,1}=19$ and $h^{1,2}=19$.

The homology of any  \dP9 surface is easily computed.
In particular, $h_{2}(\dP9, \Z)=10$ and
\begin{equation}
  H_{2}(\dP9, \Z)=\Span \{ l, e_1, e_2, \dots, e_9 \} 
  ,
  \label{4}
\end{equation}
where $l$ is the hyperplane divisor and each $e_{i}$ is an exceptional
divisor. In this basis, the fiber class of the \dP9 elliptic fibration
is given by $f=3l - \Sigma_{i=1}^{9} e_{i}$.  Furthermore, we
arbitrarily choose $e_{9}$ to be the zero section $\sigma$.  Which
\dP9 surface we are referring two, either $B_{1}$ or $B_{2}$, will be
clear from context so we will not further label their curves.

\subsection*{The Vector Bundle}

In addition to presenting the Calabi-Yau threefold $X$, it is
necessary to specify a gauge connection, indexed in a subgroup of
$E_{8}$, which satisfies the Hermitian Yang-Mills equations. It follows
from the work of Donaldson~\cite{11} and Uhlenbeck/Yau~\cite{12} that
this is equivalent to specifying a holomorphic vector bundle $V$ on
$X$ which is slope-stable. In this paper, we construct this bundle in
the following way.

First consider a line bundle
\begin{equation}
  L={\cal{O}}_{B_{1}}(e_{1}-e_{9})
  \label{5}
\end{equation}
on the surface $B_{1}$. Next, we construct a rank 3 vector bundle $W$
on $B_{2}$ as follows. Let ${\cal{C}}_{W} \in \Gamma {\cal{O}}_{B_{2}}(l+f)$
be a spectral cover and choose the line bundle ${\cal{N}}_{{\cal{C}}_{W}}
={\cal{O}}_{{\cal{C}}_{W}}$
over ${\cal{C}}_{W}$. Then $W$ is constructed from this data via the 
Fourier-Mukai transform of $({\cal{C}}_{W}, {\cal{N}}_{{\cal{C}}_{W}})$~\cite{FM},
\begin{equation}
  W=FM_{B_{2}}({\cal{O}}_{{\cal{C}}_{W}}).
  \label{6}
\end{equation}
Note that $W$ has structure group $U(3)$ with a non-trivial
determinant line bundle. Combining eqns.~\eqref{5} and~\eqref{6}, one
can construct a rank 3 vector bundle $V_{3}$ on $X$ as
\begin{equation}
  V_{3}=\pi_{1}^{*}(L) \otimes \pi_{2}^{*}(W).
  \label{7}
\end{equation}
Now define a line bundle $V_{1}$ on $X$ to be
\begin{equation}
  V_{1}=\pi_{1}^{*}(L^{-3}) \otimes \pi_{2}^{*}(\detinv W).
  \label{8}
\end{equation}
Given $V_{3}$ and $V_{1}$, we construct the requisite vector bundle $V$
on $X$ as a non-trivial extension
\begin{equation}
  0 \rightarrow V_{1} \rightarrow V \rightarrow V_{3} \rightarrow 0.
  \label{9}
\end{equation}
Note that because of our explicit choice of $V_{1}$, any such extension has
structure group $SU(4)$.

The vector bundles $V$ satisfying eq.~\eqref{9} correspond to
directions in the linear space of extensions, denoted
$\Ext^1(V_{3},V_{1})$. In terms of sheaf cohomology, this is
\begin{equation}
  \Ext^1(V_{3},V_{1})=H^1(X,V_{1} \otimes V_{3}^{*}).
  \label{10}
\end{equation}
Our construction will only be useful if this space is non-trivial.
$\Ext^1(V_{3},V_{1})$ can be computed as follows. First, using
eqns.~\eqref{7},~\eqref{8} and pushing the the cohomology down to the
projective plane $\CP^1$, one can show that
\begin{equation}
  H^1(X,V_{1} \otimes V_{3}^{*})=H^{0}\Big(\CP^1, \Hsupp(L^{-4})\cap
  \Hsupp(W^{*} \otimes \detinv W)\Big),
  \label{11}
\end{equation}
where $\Hsupp({\cal{S}})$ stands for the set of points in $\CP^1$
supporting the skyscraper sheaf $R^1\pi_{i}^{*}{\cal{S}}$. We find
that
\begin{equation}
  \Hsupp(L^{-4})=\{q_{1},q_{2},q_{3},s_{1},\dots,s_{12}\}
  \label{12}
\end{equation}
and
\begin{equation}
  \Hsupp(W^{*} \otimes \detinv W)=\{p_{1},p_{2},r_{1},\dots,r_{19}\},
  \label{13}
\end{equation}
where, a priori, the $q_{i}$, $s_{j}$, $p_{k}$ and $r_{l}$ are
independent points.  It is clear from eq.~\eqref{11} that if $\{q_{i},
s_{j}\}\cap\{p_{k}, r_{l}\}=\emptyset$ then $\Ext^1(V_{3},V_{1})$ is
trivial and there are no extensions $V$. This situation is not of
interest so, henceforth, we must always choose $L$ and $W$ so that at
least one point in $\{q_{i}, s_{j}\}$ is identified with at least one
point in $\{p_{k}, r_{l}\}$.  First assume only two points are
identified. For the purposes of this paper will always assume that
$s_{1}=r_{1}$. Then, it follows from eq.~\eqref{11} that
\begin{itemize}
\item \itemlefteq{$s_{1}=r_{1}$:}
  \begin{itemequation}
    \Ext^1(V_{3},V_{1})=\C,
    \label{14}
  \end{itemequation}
\end{itemize}
where $\C$ is the complex numbers. Next let us assume that,
in addition to $s_{1}=r_{1}$, a second and third pair of points are
identified, which we choose to be $q_{1}=p_{1}$ and $q_{2}=p_{2}$
respectively. Then
\begin{itemize}
\item \itemlefteq{$s_{1}=r_{1}$, $q_{1}=p_{1}$:} 
  \begin{itemequation}
    \Ext^1(V_{3},V_{1})=\C^{\oplus2}
    \label{15}
  \end{itemequation}
\end{itemize}
and 
\begin{itemize}
\item \itemlefteq{$s_{1}=r_{1}$, $q_{1}=p_{1}$, $q_{2}=p_{2}$:}
  \begin{itemequation}
    \Ext^1(V_{3},V_{1})=\C^{\oplus3}.
    \label{16}
  \end{itemequation}
\end{itemize}
Three pairs of points, each pair consisting of one point from the base
of $B_{1}$ and one from the base of $B_{2}$, can always be identified
without restricting the moduli space. It follows that non-trivial
extension spaces of the form eqns.~\eqref{14},~\eqref{15}
and~\eqref{16} always exist. Hence, non-trivial holomorphic vector
bundles $V$ with structure group $SU(4)$ over $X$ can be constructed.
The final step is to demonstrate that these bundles are slope-stable.
Although this is not difficult to show in this case, the proof
requires a longer discussion than is suitable for this paper. Here we
simply state that these holomorphic bundles are indeed slope-stable.
The complete proof will be presented in~\cite{13}.

Given eqns.~\eqref{7},~\eqref{8} and~\eqref{9}, it is straightforward
to compute the Chern classes of $V$. Relevant to this paper is the
fact that $c_{1}(V)=0$ and $c_3(V)=0$. Compactifying the hidden sector
of the $E_{8} \times E_{8}$ heterotic string on the Calabi-Yau
threefold $X$ with structure group $SU(4)$ vector bundle $V$ will lead
to an $N=1$ supersymmetric $Spin(10)$ gauge theory in four-dimensions.
It remains to compute the low energy matter spectrum of this theory.

\subsection*{The Spectrum}

With respect to $SU(4) \times Spin(10)$ the $\Rep{248}$
representation of the of the hidden sector $E_{8}$ gauge group
decomposes as
\begin{equation}
  \Rep{248}=(\Rep{1},\Rep{45}) \oplus (\Rep{15},\Rep{1}) \oplus 
  (\Rep{4},\Rep{16}) \oplus(\barRep{4},\barRep{16}) \oplus 
  (\Rep{6},\Rep{10}). 
  \label{18}
\end{equation}
The $(\Rep{1},\Rep{45})$ are the gauginos of $Spin(10)$, the
$(\Rep{15},\Rep{1})$ corresponds to the vector bundle moduli and the
remaining representations are the matter fields.  The number of
$\Rep{45}$ representations in the low-energy theory is given by
$h^{0}(X, {\cal{O}}_{X}) =1$. Let $\Rep{R}$ be any of the remaining
representation of $Spin(10)$ in eq.~\eqref{18} and denote the
corresponding $V$ bundle by $U_{R}(V)$. As discussed in~\cite{something}, the
multiplicity of representation $\Rep{R}$ in the low-energy theory is
given by $n_{R}=h^1\big(X,U_{R}(V)\big)$.  The number of vector bundle
moduli is $n_{1}=H^1\big(X,\ad(V)\big)$, which can be computed using
the methods presented in~\cite{13A}. Here, it suffices to say that
$n_{1}>0$.

Next we turn to the $\Rep{16}$ and $\barRep{16}$ representations.
Using the fact that $c_{3}(V)=0$, the Atiyah-Singer index theorem
tells us that $n_{16}=n_{\overline{16}}$. Furthermore, it follows from
the fact that $h^1(X,V_{1})=h^1(X,V_{3})=0$ that
\begin{equation}
  n_{16}=n_{\overline{16}}=0.
  \label{19}
\end{equation}
Hence, no $\Rep{16}$ and $\barRep{16}$ representations appear in the
low-energy theory. Now consider the multiplicity of the $\Rep{10}$
representation. This is given by
\begin{equation}
  n_{10}=h^1(X, \wedge^{2} V).
  \label{20}
\end{equation}
Using eqns.~\eqref{7},~\eqref{8},~\eqref{9} and pushing the cohomology
down to the base $\CP^1$, one can show that
\begin{equation}
  h^1(X, \wedge^{2} V)=
  2h^{0}\Big(\CP^1, \Hsupp(L^{2}) \cap \Hsupp(\wedge^{2}W)\Big),
  \label{21}
\end{equation}
where
\begin{equation}
  \Hsupp(L^{2})=\{q_{1},q_{2},q_{3}\}
  \label{22}
\end{equation}
and 
\begin{equation}
  \Hsupp(\wedge^{2}W))=\{p'_{1},p'_{2},p'_{3},p'_{4},p'_{5}\}.
  \label{23}
\end{equation}
The $q_{i}$, $i=1,2,3$ appeared in eq.~\eqref{12} and $p'_{j}$,
$j=1,\dots,5$ are five new independent points in $\CP^1$. To assure
the existence of $V$, we will always take $s_{1}=r_{1}$ and, hence,
from eq.~\eqref{14} $\Ext^1(V_{3},V_{1})$ is at least $\C$.  Let
us first assume that the points $p'_{j}$ are all independent of each
other and are not identical with any point $q_{i}$.  Then it follows
from eqns.~\eqref{20} and~\eqref{21} that
\begin{itemize}
\item \itemlefteq{$p'_{j} \neq p'_{j'}$, $p'_{j} \neq q_{i}$:}  
  \begin{itemequation}
    n_{10}=0.
    \label{24}
  \end{itemequation}
\end{itemize}
Let us now assume that one point $p'_{j}$ is identical with one point
$q_{i}$.  Without loss of generality we can choose $p'_{1}=q_{1}$.
Then
\begin{itemize}
\item \itemlefteq{$p'_{1}=q_{1}$:}
  \begin{itemequation}
    n_{10}=2.
    \label{25}
  \end{itemequation}
\end{itemize}
Now take two independent points in $p'_{j}$ and identify each with a
different point in $q_{i}$.  We can, without loss of generality,
choose $p'_{1}=q_{1}$ and $p'_{2}=q_{2}$. Then from eqns.~\eqref{20}
and~\eqref{21} we find
\begin{itemize}
\item \itemlefteq{$p'_{1}=q_{1}$, $p'_{1}=q_{1}$:}
  \begin{itemequation}
    n_{10}=4.
    \label{26}
  \end{itemequation}
\end{itemize}
Note that for all these cases $\Ext^1(V_{3},V_{1})=\C$.

We have now used all the freedom to align up to three pairs of points
on the sphere $\CP^1$. The simplest way to enlarge the number of
$\Rep{10}$ representations is as follows. Let us choose our
sub-bundles so that the $p'_{j}$ are no longer all independent. To
begin, let us arrange the bundles so that $p'_{1}=p'_{2}$, which can
be done. When this happens it turns out that this point must also be
identical to $p_{1}$. Now take $p_{1}=q_{1}$.  It follows from
eq.~\eqref{15} that, in this case, the space of extensions enlarges to
$\Ext^1(V_{3},V_{1})=\C^{\oplus2}$.  If, in addition, we take
$p'_{3}=q_{2}$, then it follows from eqns.~\eqref{20} and~\eqref{21}
that
\begin{itemize}
\item \itemlefteq{$p'_{1}=p'_{2}=p_{1}=q_{1}$, $p'_{3}=q_{2}$:}
  \begin{itemequation}
    n_{10}=6.
    \label{27}
  \end{itemequation}
\end{itemize}
Continuing in this way, let us choose $p'_{1}=p'_{2}$ and
$p'_{3}=p'_{4}$. It turns out that these points must also equal
$p_{1}$ and $p_{2}$ respectively. Now take $p_{1}=q_{1}$ and
$p_{2}=q_{2}$. We see from eq.~\eqref{16} that, in this case,
$\Ext^1(V_{3},V_{1})=\C^{\oplus3}$.  Furthermore, it follows from
eqns.~\eqref{20} and~\eqref{21} that
\begin{itemize}
\item \itemlefteq{$p'_{1}=p'_{2}=p_{1}=q_{1}$, $p'_{3}=p'_{4}=p_{2}=q_{2}$:}
  \begin{itemequation}
    \hspace{3cm}
    n_{10}=8.
    \label{28}
  \end{itemequation}
\end{itemize}
This is as far as we can go using the five points $p'_{j}$ in
eq.~\eqref{23}. However, by generalizing the spectral cover
${\cal{C}}_{W}$ to ${\cal{C}}_{W} \in \Gamma {\cal{O}}_{B_{2}}(l+nf)$
for integer $n>1$, the number of points $p'_{j}$ can be increased
arbitrarily. In this case, we can easily find vector bundle $V$ for
which $n_{10}=10, 12,\dots$ as well. It is also possible to amend the
construction of $V$ so as to obtain any odd value of $n_{10}$. This is
beyond the scope of this paper and will be discussed in~\cite{13}.

We conclude that $V$ can be chosen so that the matter spectrum of the
four-dimensional $N=1$ supersymmetric $Spin(10)$ theory contains
$\Rep{10}$ representations only and that $n_{10}$ satisfies the
requisite condition eq.~\eqref{3}. That is, $n_{10}$ can take any
value in the interval
\begin{equation}
  6 \leq n_{10} < 12.
  \label{29}
\end{equation}

\subsection*{The \Rep{10} Mass}

The above analysis proves that one can construct vector bundles $V$
whose \emph{massless} matter spectrum consists of $n_{10}$ fundamental
$\Rep{10}$ representations satisfying eq.~\eqref{29} on a
\emph{subvariety} of its moduli space. However, it follows
from eq.~\eqref{24} that as one moves off this subvariety to generic points
in moduli space these multiplets must all become massive. From the
point of view of the four-dimensional theory, this can only occur if
quadratic pairs of $\Rep{10}$ representations have non-vanishing
couplings to vector bundle moduli. This is indeed the case.
In~\cite{13} we show that non-vanishing cubic couplings of the form
\begin{equation}
  \lambda \phi \Tr \Rep{10}\, \Rep{10},
  \label{30}
\end{equation}
where $\phi$ are vector bundle moduli, occur in the superpotential.
For non-vanishing vacuum expectation values $\langle\phi\rangle$, this
leads to mass terms for each of the $n_{10}$ $\Rep{10}$
representations where
\begin{equation}
  m=\lambda \langle\phi\rangle.
  \label{31}
\end{equation}
The exact values of the parameters $\lambda$, as well as the generic
size of the moduli expectation values $\langle\phi\rangle$, are model
dependent and beyond the scope of this paper. Be that as it may, it is
clear from discussions in the literature that there is no obstruction
to choosing these so that the typical mass $m$ satisfies constraint
eq.~\eqref{1}, where $\Lambda$ is the $Spin(10)$ strong-coupling
scale.  Hence, the low-energy theory constructed here will have
non-supersymmetric, meta-stable vacua with a long life-time.

\section*{Acknowledgments}

The authors would like to thank Ken Intriligator, Tony Pantev, and
Nathan Seiberg for valuable discussions and explanations. This
research was supported in part by the Department of Physics and the
Math/Physics Research Group at the University of Pennsylvania under
cooperative research agreement DE-FG02-95ER40893 with the
U.~S.~Department of Energy and an NSF Focused Research Grant
DMS0139799 for ``The Geometry of Superstrings''. The work of E.I.B. is
supported by NSF grant PHY-0503584.


\end{document}